\documentclass[12pt]{spieman}
\usepackage{amsmath,amsfonts,amssymb}
\usepackage{graphicx}
\usepackage{setspace}
\usepackage{tocloft}
\usepackage{lineno}
\usepackage{xcolor}

\title{Probing the Transient Far-IR Sky with PRIMA}

\author[a]{David L Clements$^*$}
\author[a]{Michael Peel}
\affil[a]{Imperial College London, Physics Department, Prince Consort Road, London, UK, SW7 2AZ}

\cftpagenumbersoff{figure}
\cftpagenumbersoff{table} 
\begin{document} 
\maketitle

\begin{abstract}
The time variable far-IR/mm sky is largely unexplored. However, when PRIMA
launches, next generation ground-based CMB experiments, including Simons Observatory and CMB-S4, will be operating. These will survey large areas of the sky for transient mm sources as a byproduct of their observations,
producing regular mm-transient alerts. The first results from the current
experiments show they can detect a wide variety of mm transients ranging from Galactic stars to extragalactic sources associated with AGNs
and other energetic phenomena, and moving Solar System objects such as asteroids. These results, and theoretical predictions, indicate that future mm/submm facilities will detect
many kinds of transient, including flaring stars, protostars, GRBs, TDEs, neutron star mergers, FBOTs, and SNe (Gamma Ray Bursts, Tidal Disruption Events, Fast Blue Optical Transients and Supernovae, respectively). New classes of mm-variable may be uncovered by CMB experiments, and transient searches at other
wavelengths, such as the optical LSST survey, will produce additional
targets to followup with PRIMA. Predicted rates for extragalactic mm transients to be detected by CMB
experiments range from 10s to 1000s of events over the lifetime of these projects. CMB-S4 is most
relevant for PRIMA, producing $\sim$100 extragalactic transients per year. Galactic transients and variable sources will also be detected, but the most common Galactic transients, flaring stars, operate
on such short timescales that direct follow-up with PRIMA will not be feasible. Variable accretion rates in
forming protostars, conversely, produce long term brightness variations that will be ideal monitoring targets. The
addition of mid- and far-IR data points for all these sources can determine much about their radiation mechanisms and
underlying physics. PRIMA followup of representative examples of various mm-transient and variable sources will thus have a powerful impact on our understanding of a wide range of astrophysical phenomena.

\end{abstract}


{\noindent \footnotesize\textbf{*}David L Clements  \linkable{d.clements@imperial.ac.uk} }

\section{Introduction}

The transient sky in the far-IR is almost completely unexplored territory since the types of large sky area, multi-epoch surveys necessary for the detection of transient and variable sources have not been practicable with past generations of instruments. For example, {\em Herschel} surveys in general used only a small number of passes over their survey areas to reach their required sensitivity (e.g. \cite{2012MNRAS.424.1614O, 2010PASP..122..499E}). While a few small fields do have multi-epoch observations over a long period (e.g. the SPIRE Dark Field \cite{p24}), these are so small in area ($<$ 1 sq. deg.) that the chances of a transient occurring in them are minimal, and they have yet to be exploited for transient studies.

At longer wavelengths, in the mm/submm, the situation is somewhat better, however, thanks to a number of cosmic microwave background (CMB) experiments such as the Atacama Cosmology Telescope (ACT)\cite{2011Swetz} and the South Pole Telescope (SPT)\cite{2011Carlstrom}. These have each scanned areas of order 1000 square degrees repeatedly over year long timescales and have successfully detected both Galactic and extragalactic transient and variable sources with peak fluxes from 143 to 443 mJy (5$\sigma$ at 90 and 220 GHz respectively) for ACT \cite{2024arXiv240908429B} and $\sim$25 (5$\sigma$ at 95 and 150 GHz) for SPT \cite{2021Guns,2023Li,2024ApJ...972....6T}, though these thresholds vary somewhat depending on transient duration and scanning strategy. They also set useful upper limits on transients known at other wavelengths \cite{2024MNRAS.529.3020H}. They have also detected Solar System objects such as asteroids \cite{2022ApJ...936..173C,2024ApJ...964..138O}, which appear as transients in the survey due to their motion. However, these cannot be considered astronomical transients and so will not be discussed further in this paper. Explicit monitoring of targets of interest in the submm by, for example, the JCMT-Transient project \cite{2017ApJ...849...43H}, using the James Clerk Maxwell Telescope (JCMT)  have also detected variable \cite{2021ApJ...920..119L} and flaring \cite{2019ApJ...871...72M} sources associated with protostars in star-forming regions. Many of these sources are likely also to exhibit transient or variable emission in the far-IR  \cite{2024AJ....167...82F}, but the lack of a suitable far-IR mission overlapping with these ground-based mm/submm observations has prevented any progress in this area. A wide range of science is enabled by the addition of far- and mid-IR fluxes to the mm data that will come from the CMB experiments. These include a proper assessment of the energetics of accretion events, whether in TDEs or accreting protostars, to a better understanding of the relativistic jets produced by GRBs. The selection of PRIMA as a potential NASA far-IR probe mission means that we must now revisit the potential for far-IR observations of transient sources, especially in light of the new generation of more sensitive CMB experiments that will be operating at the time of PRIMA launch and operations.

Future generations of CMB experiments are largely concerned with the search for B-mode polarisation in order to set constraints on the physics behind inflation \cite{2019JCAP...02..056A,2022arXiv220307638C}. To achieve the necessary sensitivity to detect and characterise B-mode polarisation, sensitivities orders of magnitude better than those that were achieved by the Planck mission are required. Large detector arrays at multiple frequencies and many repeated observations of large areas of the sky are necessary to achieve this sensitivity and to control for a variety of systematic effects and foreground emission that would otherwise be dominant \cite{2024arXiv241115959N}. An example of such an experiment is the Simons Observatory,\cite{2019JCAP...02..056A} which will include at least six Small Aperture Telescopes (SATs, 0.5m diameter) and one Large Aperture Telescope (LAT, 6m diameter), all of which will operate across six bands from 27 to 280GHz, with a total of over 60,000 detectors. Of particular interest in the context of transient observations is the LAT, which will survey 40\% of the entire sky repeatedly with a reobservation cadence of $\sim$hours to days and an overall survey timescale of about 10 years. This leads to an effective sky area of about 2 million square degrees, with good sensitivity at millimetre wavelengths, making it an ideal tool for surveying the transient millimetre sky. The SO LAT should begin its science survey observations in the second quarter of 2025. The subsequent generation CMB experiment, named CMB-S4, will include many more similar telescopes providing a similar large area, high cadence survey capability reaching higher sensitivities \cite{2019arXiv190704473A}, making it an even better tool for the detection of millimetre transients. CMB-S4 should go into operation around the same time that PRIMA is expected to launch. The two projects combined will therefore allow us to characterise not only the transient millimetre sky but, with followup observations from PRIMA, the far-IR properties of these millimetre transients, and thus the first characterisation of the transient far-IR sky. The large area CCAT-prime survey has similar characteristics to the SO LAT survey but will focus on non-CMB science at somewhat higher frequencies \cite{2023ApJS..264....7C}, providing a further source of candidate mm/submm transients if it is still in operation when PRIMA launches. 

\section{Extragalactic Transients}

Long duration gamma ray bursts (LGRBs) are expected to be the dominant source of extragalactic transients detected in the millimetre by CMB experiments \cite{2022Eftekhari}, though a wide variety of sources, from supernovae (SNe) to tidal disruption events have also been detected, in most cases through targetted observations of already known transients. Despite this wide range of physical phenomena, the emission mechanism from all these sources is dominated by synchrotron processes in shock mediated outflows. The power law index of synchrotron spectra depend on the energy distribution of the emitting relativistic electrons. For highly energetic events, such as GRBs, this leads to spectral energy distributions (SED) that rise with increasing frequency  until they reach a peak at a frequency determined by the optical thickness of the emission to synchrotron self absorption and by synchrotron aging, as the emitting electrons lose energy. Determining the frequency of this peak will allow outflow velocities and the density of the surrounding interstellar medium to be measured. High frequency observations in the far-IR will allow the frequency of this peak to be better determined, providing new insights into the nature of these sources and their host environment.

\begin{figure}
\begin{center}
\includegraphics[height=12cm]{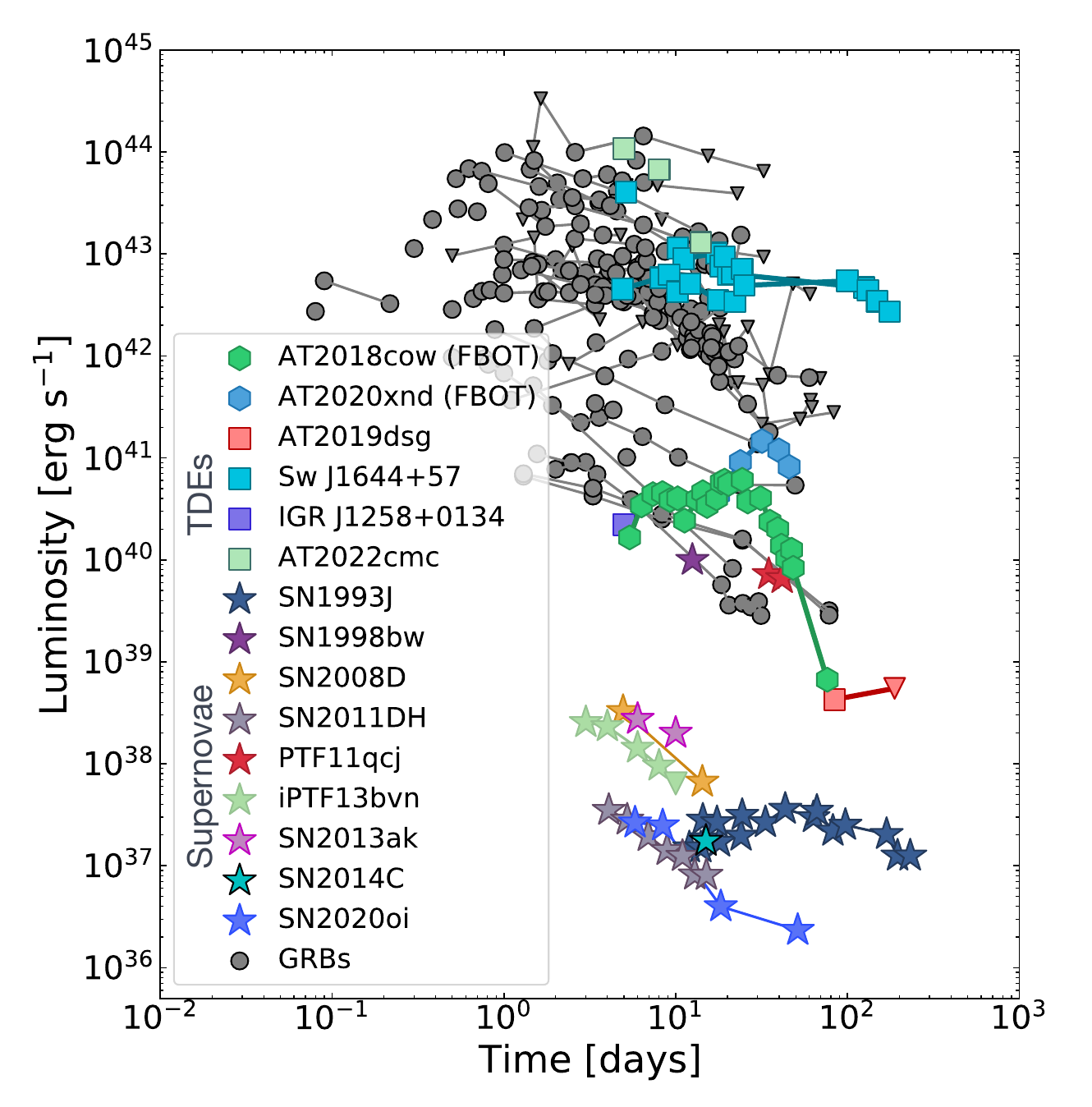}
\end{center}
\caption {Millimetre light curves of a variety of extragalactic transients, including SNe, GRBs, TDEs and FBOTs. Figure from Eftekhari et al. (2022) \cite{2022Eftekhari}.} 
\label{fig:extragalactic}
\end{figure} 

Figure \ref{fig:extragalactic}, from Eftekhari et al. (2022) \cite{2022Eftekhari} shows the millimetre light curves for a range of extragalactic transients, including GRBs, SNe, TDEs and a class of transient known as Fast Blue Optical Transients (FBOTs) whose nature is currently unclear. This shows that the transients to be found by the CMB experiments vary on timescales of days to months. The high frequency end of these SEDs is likely to vary on shorter time scales as the emitting outflow moves into less dense media and the emission becomes optically thin at lower frequencies. The TDE Swift J164449.3+573451, for example \cite{2012ApJ...748...36B} had an SED suggesting a $\sim300\mu$m flux of $\sim30$\,mJy five days after initial detection, but this falls to $\sim 6$\,mJy by 15 days while the flux at 100\,GHz stays at a roughly constant $\sim 20$\,mJy. A rapid response, on timescales of a few days after detection at 100 \,GHz, for following up this kind of source at far-IR wavelengths will thus be needed to ensure that fluxes are detectable in the long wavelength PRImager bands. As an example, pointed observations with PRImager would be able to detect the Swift J164449.3+573451 TDE with a flux at 235$\mu$m of $\sim$ 10 - 20 mJy with a 1~$\sigma$ sensitivity of 1 mJy in only a few minutes, according to the PRImager exposure time calculator, with the shorter wavelength channels providing good constraints on position of the the Synchrotron peak.

An additional class of long duration extragalactic transient where shorter wavelength observations will be useful are dust-embedded TDEs \cite{2022A&A...664A.158R, 2018Sci...361..482M}. These are systems where the X-ray and optical emission from a TDE is obscured by dust and we observe instead a light echo as the enshrouding dust heated by the TDE emission cools over time. It has been suggested that such dust enshrouded TDEs are over an order of magnitude more common than those detectable in the optical. The light echo from such events can last for many months or years and is characterised by thermal emission at temperatures starting at $\sim 1000$ K, but falling to temperatures of a few hundred K over a few thousand days, though the population is poorly characterised and there is significant variation. One source, for example, ESO 286-IG019 \cite{2022A&A...664A.158R}, seems to have a roughly constant temperature of $\sim 400$K. Very little is known about the mid- to far-IR properties of these transients since they were discovered only after the end of the Herschel mission so no observations are available. However, models\cite{2022A&A...664A.158R} suggest fluxes in the PRIMager Hyperspectral imaging bands $>$1 mJy for long periods for these dust obscured TDEs, which can be detected in less than an hour of observation according to the exposure time calculator. The PRIMA Hyperspectral imager could thus prove very useful for monitoring such objects in the long term after initial detection in the millimetre.

\section{Galactic Transients}

Transient events also occur in objects in our own Galaxy and will be detected by CMB experiments since their large footprint---for example 40\% of the entire sky for the SO LAT survey---inevitably includes the Galactic plane as well as nearby stars. Existing observations with the SPT \cite{2021Guns,2024ApJ...972....6T} and ACT \cite{2023Li} have uncovered a variety of Galactic transient sources that are associated with variable stars, M-dwarfs, binary stars and young stellar objects. These flares are usually very short in duration---minutes to hours---and have spectra that usually decline with increasing frequency, and are likely to be produced by synchrotron radiation. However, there is a class of flare, amounting to about 20\% of the events, where the SED rises with frequency whose emission may be more complex \cite{2023Li,2024ApJ...972....6T} with the higher frequency emission possibly coming from a thermal source. Studies of such objects at mid-to-far-IR wavelengths, whether during a flare on in the quiescent state, may provide important insights into the nature of these sources through determining what fraction of the emission is thermal, measuring the properties of the dust responsible for this thermal emission, and looking for signs of dust emission during the quiescent state.

\begin{figure}
\begin{center}
\includegraphics[height=9cm]{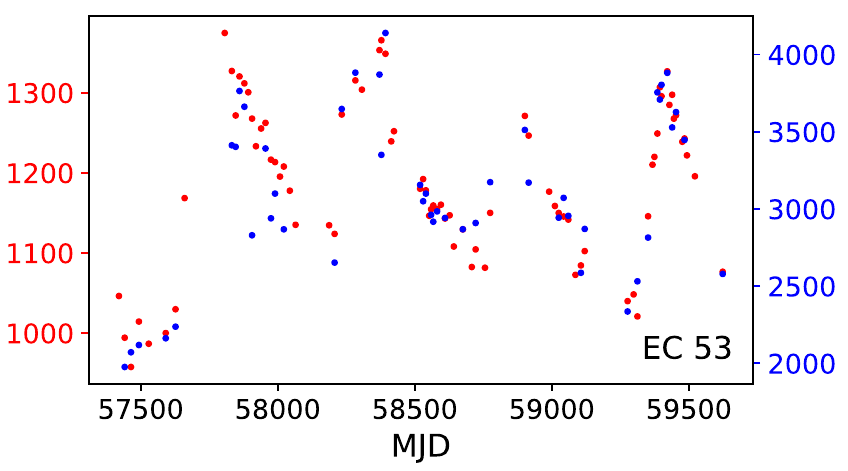}
\end{center}
\caption 
{ \label{fig:example}
Light curve of the young stellar object EC 53 in the Serpens Main star forming region as measured by the JCMT-Transient project in mJy per beam at 850 (red) and 450$\mu$m (blue) \cite{2024Mairs}. This source, which varies on day to year timescales, and others like it, have yet to have their lightcurves explored in the far-IR. From \cite{2024Mairs}.} 
\end{figure} 

An additional class of variable and/or bursting Galactic source is episodic accretion onto young stars. It has long been suspected that protostars have episodes of rapid accretion, with consequent increases in brightness, since their luminosity is lower than expected for steady accretion\cite{2012ApJ...747...52D}. The significance of such episodes in their development, however, is unclear \cite{2016ARA&A..54..135H, 2017ApJ...849...43H, 2023ASPC..534..355F}. SO and future CMB experiments will directly observe brightness variations of protostars resulting from this episodic accretion, while CCAT-Prime has some studies dedicated to monitoring appropriate fields. Previous observations at submm wavelengths using the JCMT\cite{2013ApJ...765..133J, 2017ApJ...849...43H} which observed an area of 1.6 deg$^2$ in nearby star-forming regions with a roughly monthly cadence over a 5.5 year period\cite{2024Mairs, 2021ApJ...920..119L}, found $\sim20$ protostars, about 30\% of the monitored bright sources, varying on day to year-long timescales (see Figure 2), and two sources flaring on timescales of a few days\cite{2022ApJ...937....6J, 2019ApJ...871...72M}. Meanwhile, models have shown \cite{2024AJ....167...82F} that far-IR monitoring of outbursts in protostars are the optimum way to determining the importance of episodic accretion in the mass assembly of stars. PRIMA monitoring on roughly monthly timescales of protostars found by CMB experiments, or otherwise, to be long term variables will clearly be highly valuable, while any that are found to be flaring should ideally be followed up on timescales of a few days.

\section{Other Sources of Transients}

We have so far concentrated on the kind of transients that will be detected by CMB and other experiments at millimetre wavelengths. These are of particular interest to a far-IR observatory such as PRIMA since sources that are bright in the mm are likely to be bright in the far-IR-to-submm as well, or to be such that far-IR-to-submm limits can provide insights into their nature. However, there are other sources for transient alerts that will produce potential followup targets in the far-IR that must also be considered. GRBs are an important transient population that can be detected directly in gamma rays as well as in their subsequent synchrotron emission in the mm spectrum. PRIMA followup of gamma-ray detected GRBs, currently being found at a rate of about one a day by the Fermi LAT, would be scientifically valuable in determining the synchrotron peak and in studying the interstellar medium impacted by the burst, and would yield about 40 PRIMA-accessible GRBs per year, though it is unclear what, if any, GRB detection missions will be in operation by the time PRIMA flies. There may be enhanced sensitivity instruments, capable of detecting GRBs at a higher rate and at higher redshift\cite{2024HEAD...2140603T}, or there may be no facilities at all. 

One facility that will certainly be in operation while PRIMA flies, and that will produce transient events in profusion, is the Vera Rubin Telescope with its Legacy Survey of Space and Time (VRT LSST). This will cover about 50\% of the sky and will identify optical transients at a rate of $\sim$ 10 million per night. The vast majority of these will be of little or no interest in the far-IR, but it is likely that some will be, and will provide a regular supply of potential targets of opportunity. Selection of such targets, which may include optically detected GRBs, SNe, AGN and blazar flares, will depend on the ability of LSST to classify events worthy of far-IR followup on a timeline consistent with the relevant science goals. Since VRT and the LSST will have been running for some years before PRIMA flies, the detection and classification pipelines should be well developed and fully functioning, and so will provide a regular stream of such events. Other planned future facilities, such as the 50\,m diameter Atacama Large Aperture Submillimeter Telescope (AtLAST) and CCAT-prime, may also be in operation at the same time as PRIMA and will be able to detect and supply PRIMA with transient alerts \cite{2024arXiv240413133O}.

Of more speculative interest, but potentially greater scientific value, will be searching for and following up events detected through non-electromagnetic means. Principle among these will be gravitational wave sources.  PRIMA detections would contribute to  multiwavelength analysis of any afterglows as well as allowing better characterisation of the host galaxy in the mid- to far-IR which would allow better determination of star formation rates. At the time of launch, the principal gravitational wave observatory system will be the combined LIGO-KAGRA-VIRGO array, which is detecting gravitational wave events at a rate of about two to three per day.  The key issue with gravitational wave events is the lack of localisation, but it is hoped that the VRT LSST will provide localisation for at least some fraction of these events\cite{2022Andreoni}, allowing follow-up at longer wavelengths with PRIMA. Future facilities such as the Einstein Telescope and Cosmic Explorer are expected to become operational in the mid-2030s, during PRIMA's lifetime, boosting our capacity to detect and localise gravitational wave events.

\section{Following up Mm/submm Transients with PRIMA}

Adding far-IR/submm data points to the light curves of transients, as discussed above, will clearly be useful in studying the wide variety of transient sources to be discovered by future CMB and other experiments. The scientific benefit of PRIMA can come in a number of ways, including determining the peak of synchrotron emission in non-thermal sources, revealing the dust properties such as temperature and dust mass in thermal sources, and potentially more detailed spectroscopic observations that can investigate the the detailed properties of transient sources or their immediate environment. The ability of PRIMA to follow up such sources, though, will be defined by a number of factors. Primary among these is the amount of sky accessible to the telescope at any given time - this is often referred to as the `figure of regard'. This defines whether any given transient event can be observed by the telescope at any given time. As things currently stand (Pontoppidan, private communication) the figure of regard for PRIMA is 26\%, i.e., at any given time, PRIMA can only observe 26\% of the entire sky. The event rates predicted for extragalactic transients to be detected with CMB-S4 are of order $\sim100$ per year of all types. The sky coverage of CMB-S4 will not be all sky---it is likely to be about 40\% of the sky, similar to the SO-LAT survey. The geometry of a ground-based CMB sky survey means that the astronomical positions available will cover a wide range of right ascensions (RAs), but a limited range of Declinations (Dec) - typically from small positive Dec values to some distance from the ecliptic pole. The pointing restrictions on a cold space mission at L2, such as {\em Herschel} or PRIMA, conversely, allow for a wide range of pointings perpendicular to the ecliptic, but a limited range along the ecliptic. While the details will depend on the final configuration of CMB-S4 and on the orbit actually achieved by PRIMA, it is likely that the available sky area for transient followup will be close to $0.26 \times 0.4 \approx$10\% of the sky rather than anything more favourable. There will thus be of order ten extragalactic transients detected by CMB-S4 each year that can be followed up by PRIMA. Modifications to the mission or spacecraft design that can improve the figure of regard would permit this number to increase. In this context it is worth noting that the FIRSST mission concept had an instantaneous figure of regard of 45\%. This was achieved by having the spacecraft's solar array on an articulated arm to allow the telescope a greater range of pointings while still maintaining optimal illumination of the solar panels.

The next most important factor in following up a transient event will be the speed at which the telescope can reach the target and begin observations. This is affected by two considerations. Firstly, and most obviously, is the speed at which PRIMA can slew from one position to another. In this regard PRIMA is quite fast, with a maximum slew time of 12 minutes indicated in the mission documentation. This would, in principle, allow for rapid followup of all but the briefest transient events, and will not be a significant issue for transient events lasting several hours to days. However, there is another delay that must be considered---the time taken for a transient alert to arrive and be assessed as a viable followup, and then for the appropriate commands to be sent to the telescope to initiate slewing to the target and subsequent observation. This is likely to represent a significant delay given the currently planned operating mode (Pontoppidan, private communication), timescale for transient alerts from CMB experiments ($\approx$ 30--50 hours) and previous experience with missions at L2, where the spacecraft spends a considerable time operating autonomously and only contacts the mission operations centre every few days to download data and upload new commands. Such a delay will clearly have an impact on the kind of transients that PRIMA will be able to observe. Very short duration transients, such as flaring M-stars or the reverse shocks of GRBs, operating on timescales of minutes to days, will not be accessible without significant changes in the operations scenario. However, looking at the extragalactic light curves in Figure \ref{fig:extragalactic} it can be seen that the millimetre lightcurves for a variety of types of transient, from TDEs to GRBs and SNe, and including the poorly understood FBOTs, can stay in a high state for tens of days. While spectral aging in synchrotron emission might be expected to be faster at shorter wavelengths, there is thus a good chance that PRIMA will be able to respond to many of these objects on sufficiently short timescales that they may be detectable (and, in many case, even non-detections may be useful). Most Galactic transients will in general be on shorter timescales, but there will certainly be variable and long term transient sources, such as EC 53 (see Figure \ref{fig:example}) where PRIMA can respond in a timely manner.

\section{Discussion and Conclusions}

While PRIMA has been designed to monitor regions of the sky for variability, it has not been designed to follow up transient sources. As we have discussed here it will nevertheless be able to provide mid-to-far-IR follow-up observations for reasonable samples of some classes of objects. Its utility for following up transient sources detected elsewhere could be improved in a number of ways. Firstly, if it is possible to introduce a rapid response interrupt mode, whereby the telescope can be commanded to stop any observations and move to the new transient target outside the usual schedule of telemetry uplinks, on a timescale of a few hours, then short duration transients might become viable targets. This might be achieved by a general modification to the operations scenario, or through the designation of specific periods for rapid followup during which there would be tighter coupling to ground control. Secondly, any modifications to the spacecraft that would allow a higher figure of regard would allow more transient targets to be observed. Finally, repeated observations of significant areas of the sky, if possible, might allow PRIMA to search for transients itself.

\subsection*{Disclosures}
The authors declare that there are no financial interests, commercial affiliations, or other potential conflicts of interest that could have influenced the objectivity of this research or the writing of this paper.

\subsection* {Code, Data, and Materials Availability}
All materials relevant to this paper are publicly available through the published literature.

\subsection* {Acknowledgments}
Many thanks to the UK FIR Probe and SO:UK teams for useful discussions. Thanks also to Tarraneh Eftekhari and Steve Mairs for permission to reuse their figures, and to Doug Johnstone for useful comments. This work was funded in part by STFC and UKSA.

\bibliography{report}
\bibliographystyle{spiejour}

\vspace{2ex}\noindent\textbf{David L Clements} is a Reader in Astrophysics at Imperial College London. He received his BSc and PhD from Imperial College and the University of London. He has worked on the Herschel and Planck missions and is the author of over 400 journal papers. He is also the author of the book {\em Infrared Astronomy: Seeing the Heat}. His research interests include extragalactic astronomy, observational cosmology and astrobiology, with an emphasis on far-IR and mm/submm observations.

\vspace{2ex}\noindent\textbf{Mike Peel} is a Postdoctoral Researcher at Imperial College London. He received his MPhys and PhD from the University of Manchester, and has also been a postdoc at the Universidade de São Paulo, Brasil, and the Instituto de Astroficisa de Canarias in Tenerife, Spain. His research interests span radio and submm astronomy, particularly with observations by Cosmic Microwave Background experiments. \url{https://www.mikepeel.net/}

\end{document}